\documentclass[reprint,prl,aps,showpacs,amsmath,amssymb,superscriptaddress]{revtex4-1}
\usepackage{graphicx}
\usepackage{color}
\usepackage{epstopdf}

\newcommand{\miniket}[1]{\vert#1\rangle}

\newcommand{\minibra}[1]{\langle#1\vert}

\begin{document}

\title{Experimental realisation of a delayed-choice quantum walk}

\author{Youn-Chang Jeong}
\affiliation{Department of Physics, Pohang University of Science and Technology (POSTECH), Pohang, 790-784, Korea}

\author{Carlo Di Franco}
\affiliation{Centre for Theoretical Atomic, Molecular and Optical Physics, School of Mathematics and Physics, Queen's University Belfast, BT7 1NN Belfast, United Kingdom}

\author{Hyang-Tag Lim}
\affiliation{Department of Physics, Pohang University of Science and Technology (POSTECH), Pohang, 790-784, Korea}

\author{M. S. Kim}
\affiliation{QOLS, Blackett Laboratory, Imperial College London, London SW7 2AZ, United Kingdom}

\author{Yoon-Ho Kim}
\affiliation{Department of Physics, Pohang University of Science and Technology (POSTECH), Pohang, 790-784, Korea}

%%%%%%%%%%%%%%%%%%%%

\begin{abstract}
Many paradoxes of quantum mechanics come from the fact that a quantum system can possess different features at the same time, such as in wave-particle duality or quantum superposition. In recent delayed-choice experiments, a quantum mechanical system can be observed to manifest one feature such as the wave or particle nature, depending on the final measurement setup, which is chosen after the system itself has already entered the measuring device; hence its behaviour is not predetermined. Here, we adapt this paradigmatic scheme to multi-dimensional quantum walks. In our experiment, the way in which a photon interferes with itself in a strongly non-trivial pattern depends on its polarisation, that is determined after the photon has already been detected. Multi-dimensional quantum walks are a very powerful tool for simulating the behaviour of complex quantum systems, due to their versatility. This is the first experiment realising a multi-dimensional quantum walk with a single-photon source and we present also the first experimental simulation of the Grover walk, a model that can be used to implement the Grover quantum search algorithm.
\end{abstract}

\date{\today}


\maketitle

%%%%%%%%%%%%%%%%%%%%%%%%%%%%%%%%%%%%%%%%
%%%%%%%%%%%%%%%%%%%%%%%%%%%%%%%%%%%%%%%%

The delayed-choice experiment, proposed by Wheeler~\cite{wheeler} and demonstrated in different setups~\cite{kim,peruzzodelayed,kaiser}, highlights one of the most intriguing aspects of quantum mechanics: A photon, traveling in a Mach-Zehnder interferometer, can or cannot self-interfere (and thus behave as a wave or a particle) depending on the configuration of the interferometer itself. In particular, it is possible to postpone this choice after the photon has already passed the interference stage or even after the photon has already been detected, hence the name ``delayed-choice". While, in its standard version, the choice is between having or not having interference, nothing prevents one from realising a scheme in which the choice determines different interference patterns. 
Recognising the role of quantum interference in the time evolution of quantum walks~\cite{kempe}, here we adapt the delayed-choice experiment to multi-dimensional quantum walks and experimentally demonstrate the delayed-choice two-dimensional quantum walk. This is the first experiment realising a multi-dimensional quantum walk with a single-photon source. We also present the first experimental simulation of the Grover walk, a model that can be used to implement the Grover quantum search algorithm.

The classical random walk (i.e., the mathematical description of a movement consisting of a sequence of random steps) is a fundamental process in statistical mechanics~\cite{kubo}, finding applications in various fields, from physics to computer science, from economics to biology~\cite{barber,malkiel,berg}. Interestingly, random walks are a basis for many powerful classical algorithms. The quantum walk~\cite{aharonov}, the quantum mechanical analogue of the classical random walk, behaves in a strikingly different way from its classical counterpart, due to the quantum coherences between the walker's paths. Its fast diffusive behaviour is closely related to the quadratic speedup of the Grover quantum search algorithm~\cite{grover}. Several applications of quantum walks have been found, in simulating quantum circuits~\cite{childs}, describing quantum lattice gas models~\cite{feynman}, or exploring topological phases~\cite{kitagawa,kitagawa2}. Another possible application of quantum walks is the investigation on biophysical systems, in particular the analysis and the simulation of the energy transport in the photosynthesis process~\cite{mohseni,rebentrost}.

Witnessing the strong interest in the topic, the uni-dimensional quantum walk (i.e., a quantum walk in which the movement is only allowed on a line) has already been experimentally implemented in a number of different physical systems, ranging from neutral atoms in spin-dependent optical lattices to ions in a linear ion trap, from photons in arrays of evanescently coupled waveguides to optical setups with a fibre network loop and only passive optical elements, for instance~\cite{karski,schmitz,peruzzo,schreiberintro,broome,owens,sansoni}.
And yet, the realisation of multi-dimensional quantum walks (i.e., quantum walks with movement on more than one dimension) is extremely challenging, due to the significant technological effort required. Nevertheless, the multi-dimensional scenario promises many interesting and important features that are not present in the one-dimensional case, and that clearly deserve to be observed and investigated in experiments. A noticeable example is the implementation of the quantum search algorithm, that outperforms the classical counterparts only when the dimension of the walk is higher than one~\cite{shenvi,ambanis,tulsi}. Note that one-dimensional quantum walks are not efficient for this task. A first step in experimental two-dimensional quantum walks has been recently reported in Ref.~\cite{schreiber}, where the classical interference effects of laser pulses \textit{simulate} the evolution of a quantum walker on a two-dimensional lattice.

In this paper, we demonstrate the first delayed-choice experiment applied to a two-dimensional quantum walk. In particular, we exploit the protocol proposed in Refs.~\cite{difranco,difranco2}, where it has been shown that the spatial probability distribution of the non-localised case of the Grover walk (the two-dimensional walk exploited for the quantum search) can be obtained using only a single-qubit coin and a quantum walk in alternate directions. The single-qubit coin is realised with the polarisation state of a single photon so that interference by quantum particles is responsible for the rapid spread of the walker's position in the two-dimensional lattice. The use of single photons allows us to adapt the delayed-choice scheme to the quantum walk scenario. In fact, an interesting feature of quantum walks is that the probability distribution is affected by the initial conditions, unlike their classical counterpart. In other words, the initial state of the coin (i.e., the polarisation of the single photon, in our case) determines the pattern of self-interference that gives rise to the distribution measured at the end of the walk. In our experiment, the single photon shares quantum entanglement with an ancillary photon and, because of the entanglement, its polarisation is not defined {\it a priori}. Measurement on the ancillary photon, however, has the effect of ``inducing'' a decision on which interference pattern the quantum walker has to follow. In our delayed-choice experiment, this measurement choice occurs after the photon has gone through the two-dimensional quantum walk setup and been registered at the single-photon detector.
 
 %---------------------------------------------------------------------%
\begin{figure}[t]
\includegraphics[width=3in]{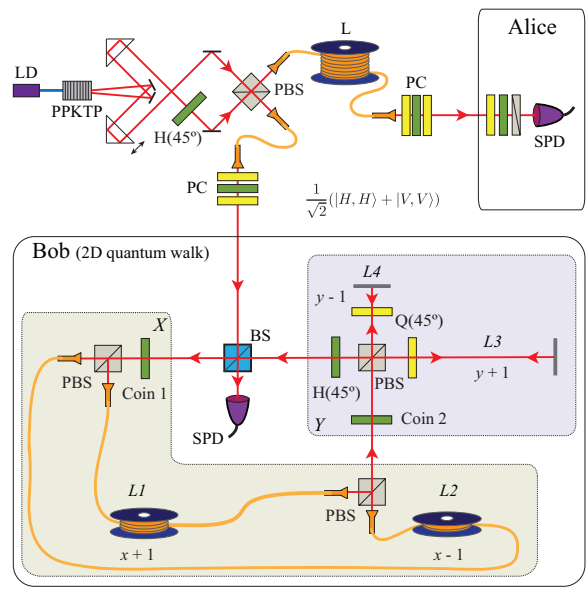}
\caption{\textbf{Experimental schematic.} The photon pair is prepared in a Bell state using the scheme in Ref.~\cite{kim03}. One photon is delayed with respect to the other by using a 340 m optical fibre spool (L). Polarisation controllers (PC) are used to ensure that the Bell state is maintained after the optical fibres. Alice can perform polarisation projection measurement in any basis with a quarter-wave plate (Q), a half-wave plate (H), and a polariser. On Bob's side, the setup implements the two-dimensional ($X$ and $Y$) quantum walk in the time domain. Coin 1 and Coin 2 perform Hadamard operations. Optical delay lines ($L1$ $\sim$ $L4$) were judiciously chosen so that the photon arrival times could be uniquely mapped to the two-dimensional lattice.}
\label{scheme}
\end{figure}
%---------------------------------------------------------------------%

Consider the experimental setup sketched in Fig.~\ref{scheme}. A photon pair is prepared via spontaneous parametric down-conversion by pumping a type-II PPKTP crystal with a 406.2 nm laser pulse. Since the coherence length of the pump laser is much shorter than the crystal length, we use the Bell-state synthesiser scheme in Ref.~\cite{kim03} to prepare a polarisation-entangled state $|\Phi^{+}\rangle=\frac{1}{\sqrt{2}}(|H,H\rangle+|V,V\rangle)$, where $|H\rangle$ and $|V\rangle$ refer to horizontal and vertical polarization, respectively. One photon is then sent to Alice via a 340 m long single-mode fibre spool (which delays the detection of the photon by approximately 1,644 ns) for delayed-choice coin-state projection and heralding the single-photon state on Bob's side. The other photon is sent to Bob to enter the two-dimensional quantum walk setup.

The two-dimensional lattice on which the walker can move is mapped to a temporal grid for the arrival times of the single photon by using the optical loop shown in Fig.~\ref{scheme}. The optical loop implements the step operation in two alternate directions $X$ and $Y$. Before each step operation ($X$ and $Y$) is taken, the coin operation (which in our experiment is chosen to be the Hadamard coin) is applied. Since the single-qubit coin is represented in the polarisation state of a single photon, the Hadamard coin operation is realised with a half-wave plate, labelled as Coin 1 for $X$ steps and Coin 2 for $Y$ steps, respectively. The $X$ step operation is implemented with a set of the polarising beam splitter (PBS) which projects the single-qubit coin into two separate output paths, two optical delay lines $L1$ and $L2$ which respectively form the walker's $X$ positions, $x+1$ and $x-1$, and the final PBS which combines the two paths into a single output mode: the walker's $X$ position corresponds to a particular time of arrival of the single photon. The $Y$ step operation is similarly implemented with different optical delay lines $L3$ and $L4$ which correspond to the walker's $Y$ positions, $y+1$ and $y-1$, respectively. The lengths ($L1, L2, L3$ and $L4$) of the optical delays are judiciously chosen in such a way that any arrival time corresponds just to a single position in the two-dimensional lattice after a specific number of steps. See Methods for details on the entangled-photon source and the 2D quantum walk setup.

After the first $X$-$Y$ step operation, the single photon returns to the 50/50 beam splitter (BS) at which there exists 50\% probability that the photon exits the optical loop to be detected at the single-photon detector (SPD). Clearly, the probability distribution for the four time grids (corresponding to a number of steps $n=1$), reconstructed from the time-correlated single-photon counting (TCSPC) events would not exhibit any interesting behaviours. The single photon that returns to the optical loop at this stage then undergoes the 2D quantum walk at $n=2$ and this result can be observed in the TCSPC measurements. Note that, in our setup, increasing the steps of the quantum walk is quite straightforward and there is no need for further resources in increasing the steps. The limiting factors are the overall loss in the optical loop and finding the proper delay lengths to make sure that time grids are experimentally identifiable. It is also interesting to note that, in our scheme, it is possible to observe all successive steps of quantum walks simultaneously as the $n+1$th step and the $n$th step are measured in the same TCSPC measurement and the former is simply delayed with respect to the latter by the total optical loop delay time. In particular, the latest arrival time corresponding to a $n=4$ two-dimensional lattice site is delayed by 535.4 ns with respect to the $n=0$ event. Since Alice's state projection and photon detection occur much later than those of the loop delays in the 2D quantum walk setup (we observe up to $n=4$ events), our experiment indeed satisfies the delayed-choice condition, that is, the `choice' is postponed until after the detection of the quantum-walk photon. Details on the data acquisition and analysis can be found in the Methods section.

%---------------------------------------------------------------------%
\begin{figure}[t]
\includegraphics[width=3in]{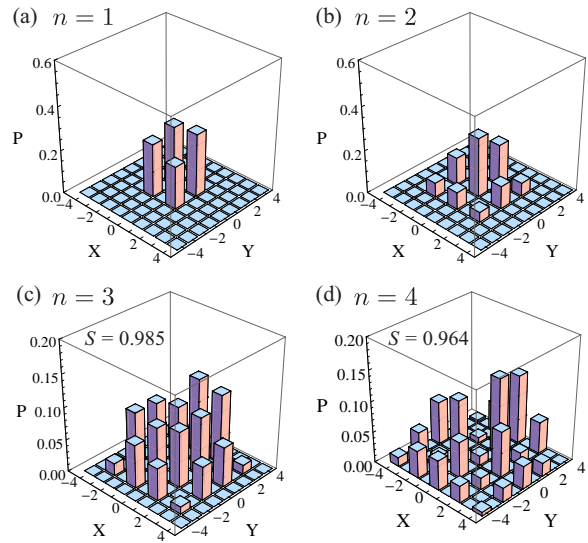}
\caption{\textbf{Evolution of probabilities in the delayed-choice 2D quantum walk.} The delayed-choice coin state is $|L\rangle$, corresponding to Alice's projection on $|R\rangle$. No quantum feature is observed up to the number of steps $n=2$, for which the theoretical quantum and classical distributions are the same. From $n=3$, quick spread out of the probability distribution, typical of the 2D quantum walk, is clearly visible. This fast diffusive behaviour is due to the quantum coherences between the walker's paths. The asymmetry in the experimental data comes from the fact that $X$-$Y$ step operations are slightly imbalanced, due to different losses. $S$ represents the similarity between the theoretical and the experimental probability distributions.}
\label{inputL}
\end{figure}
%---------------------------------------------------------------------%

%%%%%%%%%%%%%%%%%%%%%%%%%%%%%%%%%
%\section{Two-dimensional quantum walks}

Figure \ref{inputL} shows the experimentally obtained probability distributions for the 2D quantum walk with the single photon. The coin state is chosen to be $|L\rangle$ by Alice's $|R\rangle$ projection measurement in the delayed-choice setting. The experimental data clearly show that the probability distribution quickly spreads out and this is particularly evident after $n=3$. For instance, at $n=4$, the classical walker's mean position would be $(0,0)$ and our quantum walker's mean position is calculated to be $(-0.30, 0.41)$ which is quite close to the classical walker's mean position. However, we start to see something interesting if we look at the spread of the probability distribution $\{ \mathcal{P}(i,j) \}$ measured with the variance defined as $v=\sum_{i,j}\mathcal{P}(i,j)|r_{i,j}-\mu|^2$, where $r_{i,j}$ is the position on the lattice, and $\mu$ is the mean position, i.e., $\mu=\sum_{i,j} \mathcal{P}(i,j) r_{i,j}$. The quantum walker's variance from Fig.~\ref{inputL}(d) is 12.36 while the classical walker's variance, which can be evaluated easily from classical probability, would be 8. Note that the quantum walker's mean position deviates slightly from the classical value of $(0,0)$ because $X$-$Y$ step operations are somewhat imbalanced in the experiment due to different losses.

%---------------------------------------------------------------------%
\begin{figure}[t]
\includegraphics[width=3in]{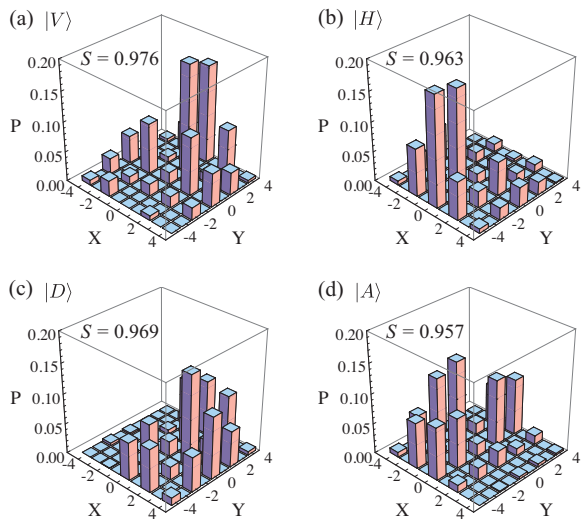}
\caption{\textbf{2D quantum walk at $n=4$ for different delayed-choice coin states.}
Alice's projection of her photon onto $|V\rangle$, $|H\rangle$, $|D\rangle$, and $|A\rangle$ results the same delayed-choice coin states. It is clear that, in quantum walks, the probability distribution is affected by the delayed-choice coin states, unlike their classical counterpart. In fact, there is a correspondence between the delayed-choice coin state and the particular direction of the $X$-$Y$ plane in which the probability is enhanced~\cite{difranco2}.}
\label{inputVHDA}
\end{figure}
%---------------------------------------------------------------------%

The theoretical probability distribution $\{\mathcal{P}_{t}(i,j)\}$ and experimental probability distribution $\{\mathcal{P}_{e}(i,j)\}$ can be compared and quantified by evaluating the quantity defined as similarity $S(\{\mathcal{P}_{t}(i,j)\},\{\mathcal{P}_{e}(i,j) \})=\left( \sum^{n}_{i,j=1} \sqrt{\mathcal{P}_{t}(i,j) \mathcal{P}_{e}(i,j)} \right)^{2}$. The similarity between the theoretical probability distributions (not shown in Fig.~\ref{inputL}) and the experimentally obtained probability distributions are displayed in Fig.~\ref{inputL}(c) and Fig.~\ref{inputL}(d) where the fast diffusive behaviour of 2D quantum walks is demonstrated.

Another unique feature of quantum walks is that the probability distribution is affected by the coin state (whether it was determined \textit{a priori} or \textit{a posteriori} does not matter). In Fig.~\ref{inputVHDA}, we show the probability distributions of the 2D quantum walk after $n=4$ steps for different delayed-choice coin states and it is straightforward to see that the coin state, which is determined remotely after the photon has already been registered at the detector, strongly affects the probability distribution.

%---------------------------------------------------------------------%
\begin{figure}[t]
\includegraphics[width=3in]{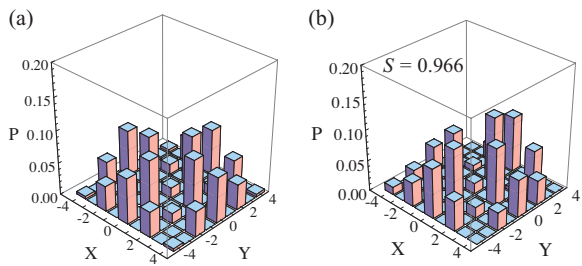}
\caption{\textbf{Delayed-choice 2D Grover walk.} (a) Theoretical probability distribution for the ideal 2D Grover walk using a four-dimensional coin and the Grover coin operation at $n=4$. (b) Experimental simulation of the 2D Grover walk at $n=4$ with the delayed-choice coin state $|R\rangle$, corresponding to Alice's projection on $|L\rangle$.}
\label{grover}
\end{figure}
%---------------------------------------------------------------------%

%%%%%%%%%%%%%%%%%%%%%%%%%%%%%%%%%%
%\section{The Grover walk}

We now discuss an important application of our scheme, the simulation of the Grover walk. In the range of 2D quantum walks that can be obtained for different coin operations, a particular attention has been given to the Grover one~\cite{shenvi,ambanis,tulsi}. This is a particular quantum walk that has recently attracted wide interest as it can implement the 2D Grover search algorithm~\cite{shenvi,ambanis,tulsi}. In order to realise the standard Grover walk, one requires a four-dimensional coin. Moreover, if the coin is embodied by two qubits, the coin operation that one has to apply each time step is a particular entangling gate, which could be very difficult to realise. Even if, as reported in Ref.~\cite{schreiber}, Schreiber {\it et al.} were able to implement an entangling gate of a specific form, realising the one required for the Grover walk in their setting is not a trivial extension. In the Grover walk, the walker is always localised unless the coin is in the particular initial state $\frac{1}{2}(|0\rangle-|1\rangle-|2\rangle+|3\rangle)$. See Methods for details on the Grover walk.

It was shown in Refs.~\cite{difranco,difranco2} that the non-localised case of the 2D Grover walk can be simulated on a two-dimensional lattice using the Hadamard coin operation and a single-qubit coin with the initial coin state $\frac{1}{\sqrt{2}}(|0\rangle - i |1\rangle)$. This can be straightforwardly realised with our setting and, in this case, the coin state corresponds to the right circular polarisation state $|R\rangle$ which is prepared remotely by Alice's delayed-choice projection on $|L\rangle$. As shown in Fig.~\ref{grover}, the experimentally observed probability distribution agrees well with the theoretically calculated ideal 2D Grover walk using a four-dimensional coin and the Grover coin operation.

%---------------------------------------------------------------------%
\begin{figure}[t]
\includegraphics[width=3in]{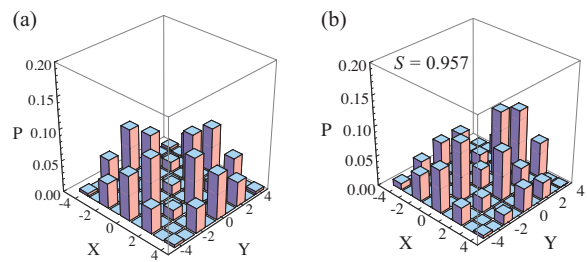}
\caption{\textbf{2D quantum walk with the maximally-mixed  coin state.} (a) Theoretical probability distribution and (b) experimentally obtained probability distribution. One may naively expect that a classical (mixed-state) coin state will not give rise to quantum behaviours, but this is not true since the quantum walk is due to the interference between the possible paths. A walk with the maximally-mixed  coin state can exhibit quantum interference as far as decoherence is negligible during the quantum walks.}
\label{mixed}
\end{figure}
%---------------------------------------------------------------------%

So far, we have considered the cases in which the coin states are prepared \textit{a posteriori} in pure quantum states. One can then ask whether a walk with the coin in a mixed state would exhibit a similar quantum behaviour. For a classical pulse, such a scenario is normally not straightforward to implement without a numerical sum of the experimental data from different settings. However, in our scheme where we make use of entangled photons and projection measurements for the coin state preparation, we can easily realise a 2D quantum walk experiment with a mixed initial coin state by simply removing the polarisation projection optical element from Alice's setup. Since Alice and Bob share a polarisation-entangled photon pair, without the polarisation projection in Alice's side, the initial coin state of Bob's photon is in a maximally-mixed state. The experimental result is shown in Fig.~\ref{mixed} and, interestingly, the 2D quantum walk with the maximally-mixed initial coin state shows quantum behaviours. In particular, the probability distribution obtained in this case is the same as the one when the initial state of the coin is pure and equal to $|L\rangle$ or $|R\rangle$~\cite{difranco2}.

In conclusion, we report the first realisation of delayed-choice experiment in a quantum walk scenario. A striking aspect of quantum mechanics is that the measurement apparatus itself determines which feature the quantum system under investigation manifests. In other words, the feature of the quantum particle which will be manifested at the measurement is not predetermined. In our delayed-choice experiment with quantum walks, the self-interference pattern depends on the polarisation of the photon embodying the walker. This polarisation is decided, due to quantum correlations, by a successive measurement on an ancillary photon. The time delay between the actual interference stage (and photon detection) and the ancilla measurement ensures that the pattern cannot be predetermined. The experimental results clearly show the characteristics of quantum walks and, with the use of single photons, we have directly demonstrated 2D quantum walks due to quantum interference, rather than simulating the effects based on wave mechanics~\cite{jeong}. We have also experimentally shown the correspondence between the Grover walk and alternate walk with a single-qubit coin. Our scheme can efficiently be expanded to implement 2D quantum walks on the order of hundreds of steps and demonstration of such massive quantum parallelism might bring out their real-life applications. Furthermore, the scheme presented in this paper can be slightly modified in order to demonstrate higher-dimensional quantum walks~\cite{roldan}, a very striking result for the simulation of multi-particle evolution and generation of genuine multi-partite entanglement. As specified in Ref.~\cite{schreiber}, the increasing of the dimension of the lattice on which the walker can move in a standard multi-dimensional quantum walk requires the exploitation of additional degrees of freedom. The reason is that the Hilbert space of the coin grows as well (and exponentially with the number of dimensions~\cite{mackay}). However, the coin remains a single qubit in our scheme, independently of the dimension of the position lattice. This is clearly an advantage in terms of experimental resources, as the only change that has to be done is an additional stage (like the ones denoted as $X$ and $Y$ in Fig.~\ref{scheme}) for each extra dimension. Therefore, the required number of optical elements scales linearly. This reflects the novel aspect of our scheme, not just to the time-multiplexing approach or the use of a heralded single-photon source.

%%%%%%%%%%%%%%%%%%%%%%%%%%%%%%%%%%%%%%%%
%%%%%%%%%%%%%%%%%%%%%%%%%%%%%%%%%%%%%%%%

\section{Methods}

\textbf{Entangled photon source.} The pump laser is a diode laser operating at 406.2 nm with the full width at half maximum (FWHM) bandwidth of 1.1 nm. The laser has 120 ns FWHM pulse width and operates at the repetition rate of 1.25 MHz. The average power of the pump laser is measured to be 10 mW. The type-II PPKTP crystal has the dimension of $1 \times 2 \times 10$ mm$^3$ and is phase-matched at 78.8$^\circ$C for degenerate (812.4 nm) and non-collinear ($\pm 1.4^\circ$ with respect to the pump direction) photon-pair generation via spontaneous parametric down-conversion. To prepare a pure polarisation-entangled photon pair via type-II SPDC, it is essential that the coherence time of the pump must be much bigger than the dimension of the nonlinear crystal. Since this condition cannot be satisfied in our case, we made use of the Bell-state synthesiser scheme in Ref.~\cite{kim03} to prepare a polarisation-entangled state $|\Phi^{+}\rangle=\frac{1}{\sqrt{2}}(|H,H\rangle+|V,V\rangle) = \frac{1}{\sqrt{2}}(|D,D\rangle+|A,A\rangle) = \frac{1}{\sqrt{2}}(|L,R\rangle+|R,L\rangle)$ where $|D\rangle\equiv (|H\rangle + |V\rangle)/\sqrt{2}$, $|A\rangle\equiv (|H\rangle - |V\rangle)/\sqrt{2}$, $|L\rangle\equiv (|H\rangle + i |V\rangle)/\sqrt{2}$ and $|R\rangle\equiv (|H\rangle - i |V\rangle)/\sqrt{2}$. This scheme allows us to generate high-quality polarisation-entangled photon pairs regardless of crystal thickness, pump bandwidth, and spectral filtering. Quantum state tomography performed on the experimental two-photon state exhibits indeed high-quality two-photon polarisation entanglement with concurrence of 0.942 and fidelity (compared to the ideal $|\Phi^{+}\rangle$ state) of 0.960.\\

\textbf{Quantum walk step operation.} For the $X$ step operation, $L1 \approx 127.8$ ns and $L2 \approx 107.2$ ns. Therefore, at the output of the second PBS, the temporal separation is 20.6 ns. For the $Y$ step operation, $L3 \approx 4.7$ ns and $L4 \approx 0.6$ ns so the effective temporal separation is $4.1$ ns. The common path $L_c$ in the optical loop is 1.3 ns. After completing the first $X$-$Y$ step operation $(n=1$), the walker's position corresponds to one of the four $(x,y)$ time grids: $(-1,-1), (-1,+1), (+1,-1)$ and $(+1,+1)$. At the second step $(n=2)$, each of the temporal modes from the first step branches into four different modes, leading to nine distinct temporal modes for the 2D quantum walk. After the $n$th step, the total number of distinct temporal modes is $(n+1)^2$.\\

\textbf{Efficiency of the setup.} The probability that a single photon at the input of the $X$ step operation (i.e., just before Coin 1 in Fig.~\ref{scheme}) is found at SPD is roughly $\eta_{det} \times \eta_{cycle}$ where the detector efficiency at 812 nm is roughly $\eta_{det}= 0.5$ and the per cycle transmittance is $\eta_{cycle}=0.207$ which includes the splitting ratio of the BS, the fibre attenuation coefficient (3 dB/km), the fibre coupling efficiency, and the FC connector loss. The present setup has rather high loss because the experimental setup was built using off-the-shelf components. Thus, there is plenty of room for improvement, for instance, by optimizing the splitting ratio of the BS, and by using low-loss fibre couplers and FC connectors.\\

\textbf{Data acquisition and analysis.} To reconstruct the 2D quantum walk probability distributions from the events distributed in the time grid, it is necessary to measure the histogram of time-of-arrival events for the single photons. This was done by observing the coincidence events between Alice's and Bob's detectors (Perkin-Elmer AQR) using a time-correlated single-photon counting (TCSPC) device (Picoharp 300 at 8 ps per bin resolution). A coincidence peak in the time grid typically has the width of 0.6 ns at FWHM so to ensure that no coincidence peaks are overlapped, we have judiciously chosen the optical delays ($L1$ $\sim$ $L4$) such that neighbouring peaks are at least 4.1 ns separated. To extract the data, a 2 ns window was defined for each time grid and the total coincidence counts within the 2 ns window was recorded. The raw coincidence counts were then corrected for accidental counts and losses. The 2D quantum walk probability distribution $P(x,y)$ for the time grid $(x,y)$ was obtained by normalising the corrected coincidence counts with the total coincidence counts for a particular step $n$. Typical data accumulation time was 5 hours per probability distribution.\\

\textbf{Grover walk.} Let us denote the Hilbert spaces of the coin subsystem and the walker subsystem as ${\cal H}_C$ and ${\cal H}_W$, respectively, and we choose $\{\miniket{x,y}_W\}$ as a basis of ${\cal H}_W$. Differently from the scheme exploited in this paper, where the walks along the horizontal axis and the vertical axis are alternated, the standard Grover walk has full freedom at each point of the space to make vertical (up/down) and horizontal (left/right) moves. Reflecting these four possibilities, the coin operation is defined in a four-dimensional space. The single time step is a sequence of the coin operation (acting only on the coin subspace)
\begin{equation}
\hat{G} = \frac{1}{2}\left(\begin{array}{cccc}
-1&1&1&1\\
1&-1&1&1\\
1&1&-1&1\\
1&1&1&-1
\end{array}
\right)
\end{equation}
and the conditional shift operation
\begin{widetext}
\begin{equation}
\begin{split}
\hat{S}=&\miniket{0}_C\minibra{0}\otimes\sum_{i,j\in \mathbb{Z}}\miniket{i-1,j-1}_W\minibra{i,j}+\miniket{1}_C\minibra{1}\otimes\sum_{i,j\in \mathbb{Z}}\miniket{i-1,j+1}_W\minibra{i,j}+\\
&\miniket{2}_C\minibra{2}\otimes\sum_{i,j\in \mathbb{Z}}\miniket{i+1,j-1}_W\minibra{i,j}+\miniket{3}_C\minibra{3}\otimes\sum_{i,j\in \mathbb{Z}}\miniket{i+1,j+1}_W\minibra{i,j}.
\end{split}
\end{equation}
\end{widetext}
Notice that, if the coin is embodied by a pair of qubits, $\hat{G}$ is an entangling gate. This standard Grover walk is interesting, in particular, because one can exploit this scheme for the implementation of the Grover search algorithm~\cite{grover}. Moreover, it has been proven that, for almost all the possible initial conditions of the coin (with the walker initially starting at the origin), one can observe a striking phenomenon of localisation~\cite{localization}. This means that the probability of finding the walker at the origin does not go asymptotically to zero for a number of time steps going to infinity.

In our scheme, the coin is embodied by a two-level system, with the coin operation given by the Hadamard gate
\begin{equation}
\hat{H}=\frac{1}{\sqrt{2}}
\begin{pmatrix}
1&1\\
1&-1
\end{pmatrix}\!.
\end{equation}
In this case, there are two different conditional shift operations,
\begin{widetext}
\begin{equation}
\hat{S}_x=\miniket{0}_C\minibra{0}\otimes\sum_{i,j\in \mathbb{Z}}\miniket{i-1,j}_W\minibra{i,j}+\miniket{1}_C\minibra{1}\otimes\sum_{i,j\in \mathbb{Z}}\miniket{i+1,j}_W\minibra{i,j}
\end{equation}
\end{widetext}
and
\begin{widetext}
\begin{equation}
\hat{S}_y=\miniket{0}_C\minibra{0}\otimes\sum_{i,j\in \mathbb{Z}}\miniket{i,j-1}_W\minibra{i,j}+\miniket{1}_C\minibra{1}\otimes\sum_{i,j\in \mathbb{Z}}\miniket{i,j+1}_W\minibra{i,j}.
\end{equation}
\end{widetext}
The evolution of the system is an alternate sequence of a coin operation followed by $\hat{S}_x$ and a coin operation followed by $\hat{S}_y$. It has been shown that this alternate walk is able to perfectly reproduce the spatial probability distribution of the non-localised case of the Grover walk~\cite{difranco,difranco2}.

%%%%%%%%%%

%%%%%%%%%%%%%%%%%%%%%%%%%%%%%%%%%

\newpage

\section{Acknowledgements}

This work was partially supported by the National Research Foundation of Korea (2011-0021452 and 2012002588). C.D.F. was supported by Basque Government grant IT472-10 and the UK EPSRC grant EP/G004579/1. H.T.L. was supported by the National Junior Research Fellowship (2012-000642). M.S.K. was supported by the NPRP 4-554-1-084 from Qatar National Research Fund. \\

%%%%%%%%%%%%%%%%%%%%%%%%%%%%%%%%%

%%%%%%%%%%%%%%%%%%%%%%%%%%%%%%%%%

\section{Author Contributions}

C.D.F., M.S.K., and Y-H.K. conceived the research, which was further developed by Y-C.J. and all co-authors. Y-C.J. and H-T.L. performed the experiment under the supervision of Y-H.K. All authors wrote the manuscript.

%%%%%%%%%%%%%%%%%%%%%%%%%%%%%%%%%

%%%%%%%%%%%%%%%%%%%%%%%%%%%%%%%%%

\section{Additional information}

Correspondence and requests for materials should be addressed to M.S.K. (m.kim@imperial.ac.uk) or Y-H.K. (yoonho72@gmail.com).

%%%%%%%%%%%%%%%%%%%%%%%%%%%%%%%%%

\end{document}